\documentclass{ws-rv9x6}
\usepackage{subfigure}   
\usepackage{ws-rv-thm}   
\usepackage{ws-rv-van}   
\makeindex
\begin{document}

\chapter[
]{Hidden Local Symmetry and Beyond
}
\label{ra_ch1}
\author{
Koichi Yamawaki
}
\address{Kobayashi-Maskawa Institute for the Origin of Particles and the Universe (KMI), 
Nagoya University,\footnote{Professor Emeritus
} \\Chikusa-ku, Nagoya, 464-8602, Japan\\ yamawaki@kmi.nagoya-u.ac.jp}

\begin{abstract}
Gerry Brown was a godfather of  
our hidden local symmetry (HLS) for the vector meson from the birth of the theory throughout his life. The HLS is originated from very nature of the nonlinear realization 
of the symmetry $G$ based on the manifold $G/H$, and thus is universal to any physics based on the nonlinear realization. Here I focus on the Higgs Lagrangian of the Standard Model (SM), which is shown to be equivalent to the nonlinear sigma model based on  $G/H= SU(2)_L\times SU(2)_R/SU(2)_{V}$ with additional symmetry, the nonlinearly realized scale symmetry. 
Then the SM does have a dynamical gauge boson of the $SU(2)_{V}$ HLS,  ``SM $\rho$ meson'', in addition to the 
Higgs as a pseudo dilaton  as well as  the NG bosons to be absorbed into the $W$ and $Z$.
Based on the recent work done with S. Matsuzaki and H. Ohki, I discuss 
a novel possibility that the SM $\rho$ meson acquires kinetic term by  the SM dynamics itself, which then stabilizes
the skyrmion dormant in the SM as
a viable candidate for the dark matter, what we call  ``Dark SM skyrmion (DSMS)''. 
\end{abstract}
\newpage
\body

\section{Introduction
}\label{ra_sec1}

HLS song to Gerry:\\
{\bf H}earty thanks to you Gerry picking me the blueberry\\
{\bf L}ong time passing a summer at garden of Brown Gerry \\
{\bf S}weet and sour memory everlasting with a smile of Grand Gerry \\

Gerry Brown was enthusiastic about the Hidden Local Symmetry (HLS) \cite{Bando:1984ej,Bando:1987br} and the associated Vector Manifestation (VM) of chiral symmetry \cite{Harada:2000kb,Harada:2003jx}, probably even more than myself and my HLS collaborators \cite{Brown:2001nh}. 
He as an editor of the Physics Reports encouraged (sometimes urged) us to publish the HLS review articles in Physics Reports, the result being  
one at tree level \cite{Bando:1987br} and another at loop level \cite{Harada:2003jx} .
In contrast to Gerry who was  working on the hadron physics as well as nuclear physics, my main interest was not the hadron physics itself but the physics hidden behind  the 
standard model (SM) Higgs, although I mostly worked on  the concrete realization of the HLS in the hadron physics. HLS in fact proved very 
successful in describing the experimental facts in hadron physics, which then  could be an excellent monitor of any theory
giving rise to the nonlinear realization, or the spontaneous breakdown of the symmetry.  

The HLS as it stands is very universal,  not just in hadron physics and particle physics, but in any system described by the nonlinear realization of the symmetry
$G$ spontaneously broken down to $H (\in G)$. 
The
nonlinear sigma model based on $G/H$ is always gauge equivalent to the HLS model having a larger symmetry $G_{\rm global} \times H_{\rm local}$, the $H_{\rm local}$
being the HLS which is a spontaneously broken gauge symmetry. 
 At classical level the gauge boson of the HLS is merely an
auxiliary field as a static composite of the Nambu-Goldstone (NG) bosons living on $G/H$, and thus can be solved away so that the theory is reduced back to the original model without HLS.
However, the HLS gauge boson at quantum level develops the
kinetic term due to the own dynamics of  the nonlinear sigma model, or those of the underlying theory if any (See Ref.\cite{Bando:1987br,Harada:2003jx} for discussions and concrete calculations).  

This is 
analogous  to the dynamical generation of the composite scalar meson in the Nambu-Jona-Lasinio (NJL) model, where the composite scalar is introduced as the auxiliary field
at classical level and then the quantum theory develops the kinetic term as well as the quartic self coupling (corresponding to the kinetic term of the Yang-Mills field for HLS), the result being equivalent to the linear sigma model with Yukawa coupling
(corresponding to the gauge coupling to matter in the HLS theory) \cite{Eguchi:1976iz}. This method was fully developed \cite{Bardeen:1989ds}
in the  top quark condensate model  \cite{Miransky:1988xi} based on the explicit  (gauged) Nambu-Jona-Lasinio model. 

Apart from the dynamical generation of the HLS gauge boson, the concept of the HLS as the effective theory of the underlying QCD-like theories
has been developed:   a manifestation of the Seiberg duality as the magnetic gauge symmetry \cite{Harada:1999zj,Harada:2003jx}, an infinite tower of HLS (``Moose'') 
 of the deconstructed/latticized gauge theory in the extra dimensions \cite{ArkaniHamed:2001ca}, further a basis of the holographic QCD \cite{Son:2003et,Sakai:2004cn},
 and so on. 

Here I focus on  novel aspects of the SM Higgs Lagrangian with Higgs mass 125 GeV, which to our surprise has far richer physics than ever recognized, without recourse to
the UV completion \cite{Matsuzaki:2016iyq}: we have recently found \cite{Fukano:2015zua} (See also \cite{Yamawaki:2015tmu}) that the SM Higgs Lagrangian is straightforwardly rewritten in the form of the scale-invariant nonlinear sigma model similar to the effective theory \cite{Matsuzaki:2015sya} of the walking technicolor \cite{Yamawaki:1985zg}, i.e., the nonlinear sigma model for both the chiral $SU(2)_L \times SU(2)_R$ symmetry and the scale symmetry realized nonlinearly, near the ``conformal limit'' (similar to the BPS limit \cite{Harvey:1996ur}), with the quartic coupling $\lambda \rightarrow 0$ with the Higgs VEV $v=$ fixed, where the SM Higgs boson $\phi$
is nothing but a pseudo-dilaton, pseudo Nambu-Goldstone (NG) boson of the spontaneously broken scale symmetry. 
Accordingly, it has an $SU(2)_V$ HLS and its dynamical  gauge boson, ``SM rho meson'',  as an analogue of the QCD rho meson. 
The resultant theory takes the same form as the scale-invariant HLS model \cite{Kurachi:2014qma}
 considered for the waking technicolor up to the explicit scale-symmetry breaking potential.

We then have found  \cite{Matsuzaki:2016iyq} that  the SM rho meson can be 
 dynamically generated by the SM dynamics itself without recourse to the UV completion beyond the SM, 
based on the one-loop calculation \cite{Harada:2003jx} of the the kinetic term of the HLS gauge boson in the general nonlinear sigma model.  
This led us to an amazing fact \cite{Matsuzaki:2016iyq}  that the
  dark matter candidate does already exist {\it inside  but not outside}  the SM (``dark side'' of the SM) , namely 
  the dynamically generated HLS gauge boson, the SM rho meson, can stabilize  
the {\it skyrmion}, ``dark SM skyrmion (DSMS)'' denoted as $X_s$. 

This is in an analogous way to the well-known mechanism \cite{Igarashi:1985et,Park:2003sd,Ma:2013ela} that the QCD skyrmion (say, nucleon with $I=J=1/2$) is
stabilized by the HLS rho meson in QCD, up to a notable difference that  the kinetic term of the QCD rho meson
is already generated by the underlying QCD not by the nonlinear sigma model own dynamics.\cite{Zahed:1986qz} 
Here we consider a scalar DSMS with $I=J=0$ having a topological charge $Q_{X_s}=1$ of  $U(1)_{X_s}$.  
The idea to identify the skyrmion as a dark matter  in a certain generalization of the SM Higgs sector (some new physics beyond the SM)  is not new (See e.g., Ref. \cite{Kitano:2016ooc} and references cited therein.)  In contrast, it is really novel to find the dark matter candidate inside the SM as it is. 

A salient feature of the DSMS is the coupling to the Higgs as a pseudo dilaton, which  is unambiguously  determined in the low energy limit due to the low energy theorem of the scale symmetry as is well known for the pseudo dilaton in a different context \cite{Carruthers:1971vz}. Accordingly, together with the nature of the soliton extended object of strong coupling system, the DSMS yields a novel dark matter phenomenology \cite{Matsuzaki:2016iyq}. Thus the dormant {\it new physics inside the SM} awaken!

This  is 
in sharp contrast to the current view that  the dark matter
is definitely originated from the physics ``beyond the SM''.
In fact the SM Higgs Lagrangian written in the form of the {\it linear} sigma model  is
usually understood as something different from the {\it nonlinear} sigma model regarded as the strong coupling (heavy Higgs mass) limit with the Higgs decoupled,
which is in obvious 
disagreement with the reality of 125 GeV Higgs, and hence regarded  irrelevant to the HLS as well. However,  even for the light Higgs it has obviously the
 symmetry $G=SU(2)_L \times SU(2)_R$ spontaneously broken down to $H=SU(2)_{L+R}=SU(2)_V$ and thus $G$ can be realized 
 nonlinearly by the Nambu-Goldstone (NG) bosons living on the manifold $G/H$, and hence does have an $SU(2)_V$ HLS. 
 As mentioned above, such a light SM Higgs near the conformal/BPS limit  simply becomes a pseudo dilaton instead of being decoupled,  giving rise to the same nonlinear realization of $G/H$ plus the nonlinearly realized scale symmetry, the result being the scale-invariant HLS
 Lagrangian similarly to that \cite{Kurachi:2014qma} considered in the walking technicolor as mentioned above.

It is further well known in the hadron physics \cite{Igarashi:1985et} that the gauge boson of the HLS can stabilize the skyrmion, with the  kinetic term becoming
precisely equal to the Skyrme term in the heavy mass limit of the rho meson where the rho meson field behaving like the auxiliary field composed of the nonlinear pions.
It has also been shown \cite{Park:2003sd,Ma:2013ela,Kitano:2016ooc} that  including the scalar meson (corresponding to the SM
Higgs as a pseudo dilaton in our case) does not invalidate the skyrmion, but rather makes the skyrmion mass lighter, 
as was particularly discussed \cite{Park:2003sd} in essentially the same form as our scale-invariant HLS model of the SM Higgs Lagrangian
(up to the potential term).

We found \cite{Matsuzaki:2016iyq} that such a lighter mass shift  of the skyrmion due to the pseudo dilaton for DSMS is in accord with the current direct detection experimental limit LUX 2006 \cite{Akerib:2016vxi} which yields unambiguously, through the characteristic low energy theorem, the {\it upper bound (instead of lower bound)} of the mass of the DSMS to be very light $M_{X_s}\lesssim 38$ GeV.  As such a light particle, it is further constrained 
by the most stringent LHS data on the Higgs invisible decay \cite{CMS:2016rfr} as $M_{X_s} \lesssim 18$ GeV, which is contrasted with most of the
WIMP dark matter candidates having mass of order of 100 GeVs. Furthermore it is crucial that the annihilation cross section can be roughly evaluated by the {\it extended size of the soliton DSMS}, which
we estimated  in the limit of heavy SM rho mass limit (Skyrme term limit) in rough consistency with the presently observed relic abundance \cite{Ade:2013zuv}. 
\\

In the next two sections
I will recapitulate Ref.\cite{Fukano:2015zua} (see also Ref.\cite{Yamawaki:2015tmu})  showing that 
the SM Higgs Lagrangian usually written in the form of {\it linear} sigma model actually can be straightforwardly rewritten  into the {\it nonlinear}
sigma model based on $G/H$, with an additional symmetry, the scale symmetry, which is also spontaneously broken and realized nonlinearly by another (pseudo) NG boson, the (pseudo) dilaton, which is nothing but the SM Higgs with mass of 125 GeV.  Then it will be further shown to be gauge equivalent to the scale-invariant version of the HLS Lagrangian. In section 4, before discussing the dynamical generation of the HLS gauge boson a la Ref.\cite{Harada:2003jx} in section 5, 
I will discuss a well-known good example of the dynamical generation of the kinetic term of the auxiliary field, namely the composite Higgs in the NJL model 
where the auxiliary Higgs field in fact becomes dynamical at the quantum level and the
system becomes equivalent to the Higgs-Yukawa model  (linear sigma model) \cite{Eguchi:1976iz} (as to the conformal/BPS limit, see Ref.\cite{Yamawaki:2015tmu}). 
In section 6 I will  discuss the recent result \cite{Matsuzaki:2016iyq} that 
the skyrmion in the SM does exist, stabilized by the SM rho meson, and it is a viable candidate for the dark matter, DSMS, totally within the
SM without physics beyond the SM.
Section 7 is devoted to summary and discussions where some possible  UV completion of the SM such as the walking technicolor  will also be addressed in the context of the DSMS.

\section{SM Higgs as a Scale-invariant Nonlinear Sigma Model \cite{Fukano:2015zua}}
\label{sNLS}

The SM Lagrangian takes the form:
\begin{eqnarray}
{\cal L}_{\rm Higgs}&=&
 |\partial_\mu h|^2 -\mu_0^2 |h|^2 -\lambda|h|^4 
 \label{Higgs}
 \\
&=& \frac{1}{2} \left[
\left(\partial_\mu {\hat \sigma}\right)^2 +\left(\partial_\mu {\hat \pi_a}\right)^2
\right]
-\frac{1}{2}
\mu_0^2  \left[ 
{\hat \sigma}^2+{\hat \pi_a}^2 
\right]-\frac{\lambda}{4} \left[ 
{\hat \sigma}^2+{\hat \pi_a}^2 
\right]^2 
 \label{sigma}\, ,
 \end{eqnarray}
where 
we have rewritten the conventional form in Eq.(\ref{Higgs})  
into the $SU(2)_L \times SU(2)_R$ linear sigma model in Eq.(\ref{sigma})  through
\begin{equation}
h=\left(\begin{array}{c}
\phi^+\\
\phi^0\end{array}  \right)=\frac{1}{\sqrt{2}} \left(\begin{array}{c}
i{\hat \pi}_1+{\hat \pi}_2\\
\hat{\sigma}-i {\hat \pi}_3\end{array}\right)\,,
\end{equation} 
with
the potential in the form:
\begin{eqnarray}
V(\hat{\sigma},\hat{\pi})&=&\frac{1}{2}
\mu_0^2  \left[ 
{\hat \sigma}^2+{\hat \pi_a}^2 
\right]+\frac{\lambda}{4} \left[ 
{\hat \sigma}^2+{\hat \pi_a}^2 
\right]^2 = \frac{1}{2}
\mu_0^2 \sigma^2 + \frac{\lambda}{4} \sigma^4\,,\\
 \sigma^2(x) &\equiv&  {\hat \sigma}^2(x) +{\hat \pi_a}^2(x)
\end{eqnarray}
which has a minimum  at the chiral-invariant circle:
\begin{equation}
\langle \sigma^2(x)\rangle = \frac{-\mu_0^2}{\lambda}\equiv v^2=(246\,{\rm GeV})^2\,.
\label{chicircle}
\end{equation}

The SM Higgs Lagrangian can be further rewritten into
\begin{equation}
{\cal L}_{\rm Higgs}= \frac{1}{2} {\rm tr} \left( \partial_\mu M\partial^\mu M^\dagger \right)
 - \left[\frac{\mu_0^2}{2} {\rm tr}\left(M M^\dagger\right)+\frac{\lambda}{4}  \left({\rm tr}\left(M M^\dagger\right)\right)^2\right] \,,
\label{sigmaMatrix}
\end{equation}
 with the $2\times 2$ matrix $M$ 
\begin{equation}
M=(i \tau_2 h^*, h) = \frac{1}{\sqrt{2}}\left({\hat \sigma}\cdot  1_{2\times 2} +2i {\hat \pi}\right)\, \quad \left({\hat \pi} \equiv {\hat \pi}_a \frac{\tau_a}{2}\right)\,,
\end{equation}
which transforms under $G=SU(2)_L\times SU(2)_R$ as:
\begin{equation}
M \rightarrow g_L \, M\, g_R^\dagger \,,\quad \left(g_{R,L} \in SU(2)_{R,L}\right)\,.
\end{equation}

Now any complex matrix $M$ can be decomposed into the Hermitian (always diagnonalizable) matrix $H$  and unitary matrix $U$ as $M=HU$ 
( ``polar decomposition'' ) \cite{Bando:1987br} : 
 \begin{equation}
 M(x) = H(x)\cdot U(x)\,, \quad H(x)=\frac{1}{\sqrt{2}} \left(\begin{array}{cc}
 \sigma(x) & 0\\
 0  &\sigma(x)
 \end{array}\right)
 \,, \quad U(x)= \exp\left(\frac{2i \pi(x)}{F_\pi}\right)  \,,
 \label{Polar}
 \end{equation}
 with $\pi(x)=\pi^a(x) \frac{\tau^a}{2} \,(a=1,2,3)$ and $F_\pi=v=\langle  \sigma(x) \rangle$. The chiral transformation of $M$ is 
 carried by $U$,
 while $H$ is a chiral singlet such that:
 \begin{equation}
 U \rightarrow g_L \, U\, g_R^\dagger\,,\quad H \rightarrow H\,,
 \label{transformation}
 \end{equation}
where $g_{L/R} \in SU(2)_{L/R}$.
 Note that {\it the radial mode $\sigma$ is a chiral-singlet in contrast to $\hat \sigma$ which is a chiral non-singlet} transformed 
 into the chiral partner ${\hat \pi}_a$ by the chiral rotation.

 We can parametrize $\sigma(x)$  as the nonlinear base of the scale transformation:
 \begin{equation} 
 \sigma(x) =v \cdot \chi(x)\,,\quad \chi(x)=\exp\left(\frac{\phi(x)}{F_\phi}
 \right)
 \,,
 \label{NLscale}
 \end{equation}
 where $F_\phi=v$ is the decay constant of the dilaton $\phi$ as the SM Higgs. 
 \footnote{
 The scale (dilatation) transformations for these fields are 
 \begin{equation}
 \delta_D \sigma =(1 +x^\mu \partial_\mu) \sigma \,, \qquad  
\delta_D \chi=(1+x^\mu \partial_\mu) \chi\,, \qquad 
\delta_D \phi= F_\phi+x^\mu \partial_\mu\phi\,. 
 \end{equation}
Note that $\langle  \sigma(x)\rangle= v \langle \chi(x) \rangle = v\ne 0$ breaks spontaneously the scale symmetry, but not the chiral symmetry, since 
$ \sigma(x)$ ($\chi(x)$ and $\phi$ as well) is a chiral singlet.
This is a nonlinear realization of the scale symmetry: 
the $\phi(x)$ is a dilaton, NG boson of the spontaneously broken scale symmetry. Although $\chi$ is a dimensionless field,
it transforms as that of dimension 1, while $\phi$ having dimension 1 transforms as the dimension 0, instead.
The {\it physical particles are $\phi$ and $\pi$} which are
defined by the  nonlinear realization, {\it in contrast to the tachyons $\hat \sigma$ and ${\hat \pi}_a$}. 
}

 With Eqs.(\ref{Polar})  and (\ref{NLscale}), we can straightforwardly rewrite the SM Higgs Lagrangian Eq.(\ref{sigmaMatrix})    
 into a form of the nonlinear sigma model:\cite{Fukano:2015zua}
 \begin{eqnarray}
  {\cal L}_{\rm Higgs}
 &=&\left[ \frac{F_\phi^2}{2} \left(\partial_\mu \chi \right)^2+ \frac{F_\pi^2}{4}{\chi}^2\cdot {\rm tr} \left(\partial_\mu U \partial^\mu U^\dagger\right)\right]
  - V(\phi)\nonumber\\
 &=&\chi^2(x) \cdot \left[ \frac{1}{2} \left(\partial_\mu \phi\right)^2  +\frac{F_\pi^2}{4}{\rm tr} \left(\partial_\mu U \partial^\mu U^\dagger\right)\right] -V(\phi)\nonumber \,,\\
 V(\phi)&=& \frac{\lambda}{4} v^4 \left[\left(\chi^2(x) -1\right)^2-1\right]  =\frac{M_\phi^2 F_\phi^2}{8} \left[\left(\chi^2(x) -1\right)^2-1\right] \,,\nonumber \\
 F_\phi&=&F_\pi=v, \quad M_\phi^2 = 2 \lambda v^2\,.
 \label{SNLSM}
 \end{eqnarray}
This  is nothing but {\it the scale-invariant nonlinear sigma model},  
similar to the effective theory of the walking technicolor~\cite{Matsuzaki:2015sya},  apart from the form of the explicit scale-symmetry breaking 
potential: $V(\phi)=\frac{M_\phi^2 F_\phi^2}{4} \chi^4 \left(\ln\chi-\frac{1}{4}\right)$, with $F_\phi\ne F_\pi\ne v$ in general, in stead of that in Eq.(\ref{SNLSM}).

The scale symmetry is explicitly broken only by the  potential $V(\phi)$ such that $\delta_D V(\phi) = \lambda v^4\chi^2=-\theta_\mu^\mu$  whose scale dimension $d_{\theta}=2$ (originally the tachyon mass term), namely, the scale symmetry is broken only by the dimension 2 operator instead of dimension 4 in the walking technicolor arising from the trace anomaly of quantum mechanical origin:
This yields the mass of the (pseudo-)dilaton as the Higgs $M_\phi^2=2\lambda v^2$, which is in accord  with the Partially Conserved Dilatation Current (PCDC) for $\partial^\mu D_\mu=\theta_\mu^\mu$:
\begin{equation}
M_\phi^2 F_\phi^2=-\langle0|\partial^\mu D_\mu|\phi\rangle F_\phi=-d_{\theta} \langle \theta_\mu^\mu\rangle =2\lambda v^4\langle \chi^2(x) \rangle=2\lambda v^4\,,
\label{PCDC}
\end{equation}
with $F_\phi=v$, where $D_\mu$ is the dilatation current: $\langle 0|D_\mu(x) |\phi\rangle=-i q_\mu F_\phi e^{-i q x}$, or equivalently $\langle 0| \theta_{\mu\nu}|\phi(q)\rangle= F_\phi (q_\mu q_\nu- q^2 g_{\mu\nu}/3)$.  
 
Hence the SM Higgs
as it stands is a (pseudo) dilaton, with the {\it mass arising from the dimension 2 operator}
 in the potential, which vanishes for $\lambda\rightarrow 0$:
 \begin{equation}
 M_\phi^2=2\lambda v^2 
 \rightarrow 0
 \quad \left(\lambda\rightarrow 0\,, \,\,
 v=\sqrt{\frac{-\mu^2_0}{\lambda}} = 
 {\rm fixed}\,\ne 0\right)
 \label{conformallimit}
  \end{equation}
  (``conformal limit''\cite{Fukano:2015zua}, which corresponds to the so-called ``Bogomol'nyi-Prasad-Sommerfield (BPS) limit'' of 
't Hooft-Polyakov monopole in the Georgi-Glashow model, similarly to the SUSY flat direction.\cite{Harvey:1996ur}).
\footnote{
With vanishing  potential, $V(\phi) \rightarrow 0$, this limit still gives an  {\it interacting theory where the physical particles $\pi$ and $\phi$ have derivative coupling} in the same sense as in the nonlinear chiral Lagrangian Eq.(\ref{NLS}).  It should be contrasted to the triviality limit, $\lambda\rightarrow 0$ {\it without fixing $ v=\sqrt{\frac{-\mu^2_0}{\lambda}}\ne 0$}, which yields only a  free theory of  tachyons $\hat \pi$ and $\hat \sigma$. The interaction of course generates the trace anomaly of dimension 4 just like that of the walking technicolor, even in the conformal/BPS limit where the tree-level potential vanishes.
This limit should also be distinguished from the popular limit $\mu^2_0\rightarrow 0 $ with $\lambda=$fixed $\ne 0$, where the Coleman-Weinberg potential as the explicit scale symmetry breaking is generated by the trace anomaly  (dimension 4 operator) due to the quantum loop in somewhat different context. 
}
In fact the Higgs mass 125 GeV implies that the SM Higgs is in near conformal/BPS limit $\lambda\rightarrow 0$ with $v=$ fixed: 
\begin{equation}
\lambda=\frac{1}{2} \left(\frac{M_\phi}{v}\right)^2 \simeq \frac{1}{2} \left(\frac{125\,{\rm GeV}}{246\,{\rm GeV}}\right)^2 \simeq \frac{1}{8} \ll 1\,.
\end{equation}  
Note that {\it mass term of all the SM particles except the Higgs is scale-invariant}. 

By the electro-weak gauging as usual; $\partial_\mu U\Rightarrow {\cal D}_\mu U= \partial_\mu U -i g_2 W_\mu U +i g_1 U B_\mu$
 in Eq.(\ref{SNLSM}), we see that the mass term of $W/Z$ is scale-invariant thanks to the dilaton factor $\chi$, and so is the mass term of the SM fermions $f$: $g_Y  \bar f h f
 =(g_Y v/\sqrt{2}) (\chi \bar f f)$, all with the scale dimension 4.  
 This implies that {\it the couplings of the SM Higgs as a pseudo dilaton to all the SM particles are written in the scale-invariant form and thus 
 obey the low energy theorem of the scale symmetry} in perfect agreement with the experiments: 
 
 The low energy theorem for the pseudo dilaton $\phi(q_\mu)$ coupling to the canonical matter filed $X$ at  $q_\mu$$\rightarrow 0$ reads 
 \begin{eqnarray}
 g_{\phi X^\dagger X}=\frac{2M_X^2}{F_\phi},\quad g_{\phi \bar X X}= \frac{M_X}{F_\phi} \quad \left(F_\phi=v\right)\,,
  \label{LET}
    \end{eqnarray}
   for complex scalar and spin $1/2$ fermion, respectively \cite{Carruthers:1971vz},  which can also be read from
    the scale invariance of the mass term;
 \begin{eqnarray}
 M_X^2 \cdot \chi^2\, X^\dagger  X&=&M_X^2 X^\dagger X + \frac{2 M_X^2}{v} \phi X^\dagger X +\cdots\,,\nonumber \\
 M_X \cdot\chi\,  \bar X  X&=&M_X \bar X X + \frac{M_X}{v} \phi {\bar X} X +\cdots\,.
 \end{eqnarray}
 for the respective canonical field with 
 the canonical dimension (For the neutral scalar we have $g_{\phi X X}=M_X^2/F_\phi$). 
 See Re.\cite{Bando:1986bg,Matsuzaki:2012vc} for the general form of the low energy theorem of the scale symmetry including the anomalous dimension.

On the other hand, if we take the limit $\lambda \rightarrow \infty$, 
then the SM Higgs Lagrangian goes over to the  usual nonlinear sigma model {\it without scale symmetry}:  
\begin{equation}
{\cal L}_{{\rm NL}\sigma}=\frac{F_\pi^2}{4} {\rm tr} \left(\partial^\mu U\partial_\mu U^\dagger\right)\,,\quad \left(\lambda\rightarrow \infty\,, \,\,
 F_\pi=v=\sqrt{\frac{-\mu^2_0}{\lambda}} = 
 {\rm fixed}\,\ne 0\right)
 \label{NLS}
\end{equation}
where the potential is decoupled with $\chi(x)$ frozen to the minimal point $\chi(x)\equiv 1$ ($\phi(x) \equiv  \langle \phi(x) \rangle=v\ne 0$), so that the scale symmetry breaking is transferred from the potential  to the kinetic term, 
which is no longer transformed as the dimension 4 operator.  

The $\lambda \rightarrow \infty$ limit  is known to be a good effective theory (chiral perturbation theory) of the ordinary QCD which in fact lacks the scale symmetry at all,
perfectly consistent  with
the nonlinear sigma model, Eq.(\ref{NLS}). 

Absence of the scale symmetry in QCD corresponding to $\lambda \rightarrow \infty$ is also consistent with the well-known failure of the old idea to regard the lightest scalar $f_0(500)$ (``$\sigma$'') as the pseudo dilaton \cite{Carruthers:1971vz}: if it were the pseudo dilaton, the  low energy theorem of the scale symmetry in Eq.(\ref{LET}) would uniquely determine the low energy limit dilaton couplings to the matter (including massive pion) in units of the dilaton decay constant $f_{\sigma} (\geq f_\pi=93 {\rm MeV})$.
We may take the coupling ratio which is free from the unknown parameter $f_\sigma $, and see a typical case of couplings to $\pi$ and the nucleon:
\footnote{
The low energy theorem coupling $g_{\phi X^\dagger X}=2 M_X^2/F_\phi$ in Eq.(\ref{LET}) corresponds to $2 m_\pi g_{\sigma \pi\pi}$ with the conventional dimensionless coupling $g_{\sigma \pi\pi}=m_\pi/f_\sigma$ used here.
} $g_{\sigma \pi\pi}/g_{\sigma NN}=m_\pi/m_N$ for ${\cal L}= m_\pi g_{\sigma \pi\pi} \cdot
\sigma \pi^a \pi_a$, $g_{\sigma NN} \cdot
\bar N \sigma N$,  which predicts $g^2_{\sigma \pi\pi}\simeq (m_\pi/m_N)^2 g^2_{\sigma NN}\simeq 2$ for the observed value $g_{\sigma NN} \simeq 10$ (the value consistent with the
low energy theorem $g_{\sigma NN}= m_N/f_\sigma$, if $f_\sigma = f_\pi$). Then the pseudo dilaton width  would be 
 \begin{equation}
\Gamma_\sigma \simeq  3 \times \frac{ m_\pi^2 g^2_{\sigma\pi\pi}}{8\pi m_\sigma}\left(1-\frac{4m_\pi^2}{m^2_\sigma}\right)^{1/2}\sim 7-8\,\,  {\rm MeV} \,\, (m_\sigma=500-600 \, {\rm MeV})\,,
\end{equation}
which is compared with the experiment $\Gamma_{f_0} = 400 -700 \, {\rm MeV}$, roughly two orders magnitude smaller (unless the on-shell couplings are drastically distorted from the low energy limit values, though it is another symptom of the absence of the scale symmetry anyway)

{\it Thus there is no remnants of scale symmetry in the QCD in the free space (There could be some emergent scale symmetry
for the hot and/or dense QCD, however \cite{Lee:2015qsa}
).
This is in sharp contrast to the SM Higgs whose couplings to the SM particles (quark/lepton and $W/Z/\gamma$) all  respect the
low energy theorem of the scale symmetry even on the SM Higgs on-shell away from the low energy limit,  in perfect agreement with the experiments as mentioned before.}

\section{HLS in the SM Higgs Lagrangian, the ``SM Rho Meson'' \cite{Fukano:2015zua}}
\label{sHLS}

The SM Higgs Lagrangian was further shown \cite{Fukano:2015zua} to be {\it gauge equivalent} to the {\it scale-invariant} 
version \cite{Kurachi:2014qma} of the Hidden Local Symmetry (HLS) Lagrangian \cite{Bando:1984ej,Bando:1987br,
Harada:2003jx}, which contains {\it possible new vector boson}, ``SM rho'', {\it hidden behind the SM Higgs Lagrangian},
as an analogue of the QCD $\rho$ meson.
\footnote{
The s-HLS model was also discussed in a different context, ordinary QCD in medium.\cite{Lee:2015qsa}
}

As usual, the HLS can be made explicit by dividing $U(x)$ into two parts \cite{Bando:1984ej,Bando:1987br,
Harada:2003jx}:
\begin{equation}
 U(x)= \xi_L^\dagger(x) \cdot \xi_R(x)\,,
  \label{U:decomp}
\end{equation}  
where  
$\xi_{R,L}(x)$ transform under $G_{\rm global} \times H_{\rm local}$ as
\begin{eqnarray}
\xi_{R,L}(x) \rightarrow h(x) \cdot \xi_{R,L}(x) \cdot { g^\prime}_{R,L}^\dagger\,,\quad 
U(x) \rightarrow {\hat g}_L U(x) { g^\prime}_R^\dagger  \nonumber \\ 
 \left(h(x)\in H_{\rm local},\, {g^\prime}_{R,L}\in G_{\rm global} \right)\,.
\end{eqnarray}
The $H_{\rm local} $ is a gauge symmetry of group $H$ arising from the redundancy (gauge symmetry) how to divide
$U$ into two parts. 

Then we can introduce the HLS gauge boson,``SM rho'' meson,  $\rho_\mu(x)$ by covariant derivative as 
\begin{equation}
D_\mu \xi_{R,L}(x) = \partial_\mu \xi_{R,L} (x)-i \rho_\mu(x) \xi_{R,L}(x) \,,
\label{HLScovariant}
\end{equation}
which transform in the same way as $\xi_{R,L}$. Then we have two covariant objects transforming homogeneously under $H_{\rm local}$: 
  \begin{eqnarray}
   \{  {\hat \alpha}_{\mu,R,L}, 
   {\hat \alpha}_{\mu, ||, \perp}
   \}
  &\rightarrow& h(x) \cdot \{ {\hat \alpha}_{\mu, R,L},
  {\hat \alpha}_{\mu, ||,\perp}
  \}\cdot h^\dagger(x)\,,\nonumber \\ 
  {\hat \alpha}_{\mu,R,L}
  &\equiv& \frac{1}{i}D_\mu \xi_{R,L}
  \cdot \xi_{R,L}^\dagger
  =\frac{1}{i}\partial_\mu \xi_{R,L}
   \cdot \xi_{R,L}^\dagger 
 - \rho_\mu
 \,, \nonumber\\ 
 {\hat \alpha}_{\mu, ||,\perp}
 &\equiv & \frac{1}{2}\left({\hat \alpha}_{\mu,R}
 \pm  {\hat \alpha}_{\mu, L}
 \right) 
= 
\Bigg\{ \begin{array}{c}
  {\alpha}_{\mu ||}
   - \rho_\mu
  \nonumber \\  
 {\alpha}_{\mu \perp}  
 \end{array},,\\
      \end{eqnarray}
      where
      \begin{eqnarray}
   {\alpha}_{\mu ||}
    &=&  \frac{1}{2i} \left(\partial_\mu \xi_{R}
    \cdot \xi_{R}^\dagger
    + \partial_\mu \xi_{L}
     \cdot \xi_{L}^\dagger
      \right) =\frac{1}{F_\rho} \partial_\mu \check{\rho}  - \frac{i}{2 F_\pi^2}[\partial_\mu \pi,\pi] +\cdots\,,  \nonumber\\
    {\alpha}_{\mu \perp} 
    &=&\frac{1}{2i} \left(\partial_\mu \xi_{R} 
    \cdot \xi_{R}^\dagger
    - \partial_\mu \xi_{L}
     \cdot \xi_{L}^\dagger
       \right) \nonumber\\
       &=&\frac{1}{2i}\xi_L\cdot \partial_\mu U\cdot \xi_R^\dagger=\frac{1}{2i}\xi_R \partial_\mu U^\dagger\cdot \xi_L^\dagger\,,
                 \end{eqnarray}
 with   $\check{\rho}$ and $F_\rho$ being the NG boson  to be absorbed into the longitudinal $\rho_\mu$ and its  decay constant, respectively
   as introduced by $\xi_{R,L}=e^{i \check{\rho}/F_\rho} \cdot e^{\pm i \pi/v}$ (See the discussions below)\footnote{In the HLS papers \cite{Bando:1984ej,Bando:1987br,
Harada:2003jx} $\check{\rho}$ was denoted by $\sigma$. In order to avoid confusion we will use $\check{\rho}$ in this paper.} .
      
We thus have two independent invariants under the larger symmetry $G_{\rm global} \times H_{\rm local}$:
\begin{eqnarray}
\,
{\cal L}_A&=&v^2\cdot {\rm tr} {\hat \alpha}_{\perp}^2(x)=v^2  \cdot{\rm tr}{\alpha}_{\mu \perp}^2(x)=  \frac{v^2}{4}\cdot {\rm tr} \left(\partial_\mu U \partial^\mu U^\dagger\right) \,,\\
{\cal L}_V&=&v^2\cdot  {\rm tr} \,{\hat \alpha}_{\mu ||}^2(x) = v^2\cdot   {\rm tr} \, \left(\rho_\mu(x) - {\alpha}_{\mu ||}(x)\right)^2 \nonumber \,,\\
&=&v^2\cdot   {\rm tr} \, \left(\left(\rho_\mu(x) - \frac{1}{F_\rho} \partial_\mu {\check \rho}\right) -\frac{i}{2 F_\pi^2} [\partial_\mu \pi,\pi] +\cdots \right)^2
\end{eqnarray} 
where ${\cal L}_A$ is the original nonlinear sigma model on $G/H$, a part of the SM Higgs Lagrangian in the form of Eq.(\ref{SNLSM}).
Then the scale-invariant $G_{\rm global} \times H_{\rm local}$ model takes the form:
\begin{equation}
{\cal L}_{\rm s-HLS} = \chi^2(x) \cdot \left(\frac{1}{2} \left(\partial_\mu \phi\right)^2 + {\cal L}_A+ a {\cal L}_V\right)\,,
\label{SHLS}
\end{equation} 
with $a$ being an arbitrary parameter. The kinetic term of $\check \rho$ is normalized as $F_{\rho}^2=a F_\pi^2 = a v^2$.  

We now fix the gauge of HLS as $\xi_L^\dagger=\xi_R=\xi=e^{i \pi/v}$ such that $U=\xi^2$ (unitary gauge $\check{\rho}=0$). 
Then  $H_{\rm local}$ and $H_{\rm global} (\subset G_{\rm global})$ 
get simultaneously broken spontaneously (Higgs mechanism), leaving the 
diagonal subgroup $H=H_{\rm local}+H_{\rm global}$, which is nothing but the subgroup $H$ of the original $G$ of $G/H$: $H\subset G$. Then the
extended symmetry $G_{\rm global} \times H_{\rm local}$ is simply reduced back to the original nonlinear realization of $G$ on the manifold $G/H$,
both are gauge equivalent to each other. 

Thus the SM Lagrangian in the form of Eq.(\ref{SNLSM}) is gauge equivalent to
\begin{eqnarray}
{\cal L}_{\rm Higgs-HLS}&=& \chi^2(x) \cdot \left(
\frac{1}{2} \left(\partial_\mu \phi\right)^2 +
 \frac{v^2}{4}\cdot {\rm tr} \left(\partial_\mu U \partial^\mu U^\dagger \right) \right) -V(\phi) \nonumber \\
 &+&  \chi^2(x) \cdot F_\rho^2\cdot  
    {\rm tr}  \left(\rho_\mu(x) - {\alpha}_{\mu ||}(x)\right)^2  
\,, \quad\left(F_\rho^2= a v^2\right)
 \label{HiggsHLS}
 \end{eqnarray}
where the second line $\chi^2 a {\cal L}_V$ is the extra term which has the rho field $\rho_\mu$ as the auxiliary field. In the path integral language it is a Gaussian integral
which would not change the physics.  
In fact, at classical level in the absence of the kinetic term for the HLS gauge boson $\rho_\mu$, it is simply solved away to yield 
\begin{equation} 
\chi^2(x) \cdot 
F_\rho^2 \cdot   {\rm tr} \, \left(\rho_\mu(x) - {\alpha}_{\mu ||}(x)\right)^2 = 0\,,
\end{equation}
 after using the equation of motion 
$\rho_\mu={\alpha}_{\mu ||}$. 
Then ${\cal L}_{\rm Higgs-HLS}$ in Eq. (\ref{HiggsHLS})  is simply reduced back to the original SM Higgs Lagrangian ${\cal L}_{\rm Higgs}$ in nonlinear realization, Eq.(\ref{SNLSM}).

As we shall discuss later, however,  the auxiliary field $\rho_\mu$ actually acquires the kinetic term 
\begin{equation}
{\cal L}^{(\rho)}_{\rm kinetic} =- \frac{1}{2 g^2}\,  {\rm tr}\,  \rho_{\mu\nu}^2
\end{equation}
 by the quantum corrections, with $g$
being the gauge coupling of HLS.  
Then the quantum theory for the SM Higgs takes the form:
\begin{eqnarray}
{\cal L}_{\rm Higgs-HLS}= \chi^2
\cdot \left(
\frac{1}{2} \left(\partial_\mu \phi\right)^2 +
 \frac{v^2}{4}\cdot {\rm tr} \left(\partial_\mu U \partial^\mu U^\dagger \right) \right) -V(\phi) \nonumber \\
 +  \chi^2
  \cdot F_\rho^2 
 \cdot  {\rm tr} \, \left(\rho_\mu
  - {\alpha}_{\mu ||}
  \right)^2  
 - \frac{1}{2 g^2}\, {\rm tr}\, \rho_{\mu\nu}^2 +\cdots
 \,,
 \label{qHiggsHLS}
 \end{eqnarray}
where ``$\cdots$'' stands for other induced terms at quantum level.

When it happens, after rescaling the kinetic term of $\rho_\mu$, $\rho_\mu(x) \rightarrow g\, \rho_\mu(x)$ to the canonical one $-\frac{1}{2 }\, {\rm tr} \,\rho_{\mu\nu}^2$, 
the $\chi^2 \, a {\cal L}_V$ term yields  the scale-invariant mass term of $\rho_\mu$, 
\begin{eqnarray}
\chi^2\, a  {\cal L}_V&=& \chi^2
F_\rho^2 \cdot   {\rm tr} \, \left(g \rho_\mu
 - {\alpha}_{\mu,||}
 \right)^2=
 M_\rho^2 \,{\rm tr} (\rho_\mu
 )^2+ g_{\rho\pi\pi} \cdot 2i {\rm tr} \left(\rho^\mu\left[\partial_\mu \pi,\pi\right]\right) +\cdots\,,\nonumber \\
 M_\rho^2&=& g^2 F_\rho^2 = a (g \, v)^2\,,\quad g_{\rho\pi\pi}= \frac{F_\rho^2}{2 v^2}\, g=\frac{a}{2} g\,, 
 \label{rhomass}
 \end{eqnarray}
 with the mass acquired by the Higgs mechanism mentioned above, which provides the standard KSRF I relation for $a=F_\rho^2/F_\pi^2=F_\rho^2/v^2=2$ \cite{Bando:1984ej,Bando:1987br}:
\begin{equation}
 M_\rho^2=\left(\frac{4 F_\pi^2}{F_\rho^2}\right) g^2_{\rho\pi\pi}F_\pi^2=\frac{4}{a} g^2_{\rho\pi\pi} v^2\,.
 \end{equation} 
Note that {\it the HLS gauge boson acquires the
scale-invariant mass term thanks to the dilaton factor $\chi^2$}, the nonlinear realization of the scale symmetry, in sharp contrast to the {\it Higgs (dilaton)  which acquires mass only from the explicit breaking of the scale symmetry}.

The electroweak gauge bosons ($\in {\cal R}_\mu ({\cal L}_\mu)$) are introduced by extending the covariant derivative of Eq.(\ref{HLScovariant}) 
this time by gauging $G_{\rm global}$, which is {\it independent of $H_{\rm local}$} in  the HLS extension:
\begin{equation}
D_\mu \xi_{R,L}(x)\Rightarrow {\hat D}_\mu \xi_{R,L}(x)\equiv  \partial_\mu \xi_{R,L} (x)-i \rho_\mu(x) \, \xi_{R,L}(x)  +i \xi_{R,L}(x)\, {\cal R}_\mu ( {\cal L}_\mu)\,.
\label{fullcovariant}
\end{equation}
We then finally have a gauged s-HLS version of the Higgs Lagrangian (gauged-s-HLS):\footnote{
This form of the Lagrangian  is the same as that of the effective theory of the one-family ($N_F=8$) walking technicolor \cite{Kurachi:2014qma}, except for the shape of the scale-violating 
potential $V(\phi)$  which has a scale dimension 4 (trace anomaly) in the case of the walking technicolor instead of  2 of the SM Higgs case (Lagrangian mass term). 
}
 \begin{equation}
 {\cal L}^{\rm gauged}_{\rm Higgs-HLS}=  \chi^2(x)\cdot 
 \left[
\frac{1}{2} \left(\partial_\mu \phi\right)^2  + {\hat {\cal L}}_A+ a {\hat {\cal L}}_V
  \right] - V(\phi) + {\cal L}_{\rm kinetic}^{(\rho, {\cal L},\cal{R})} +\cdots \,, 
  \label{gaugedsHLS}
  \end{equation}
 with 
 \begin{equation}
  {\hat {\cal L}}_{A,V}= {\cal L}_{A,V} \left( D_\mu \xi_{R,L}(x) \Rightarrow {\hat D}_\mu \xi_{R,L}(x)\right) \,.
     \end{equation}
This yields, besides Eq.(\ref{rhomass}),  a notable $a-$independent relation (KSRF II) between the $\rho-\gamma$ mixing strength $g_\rho$ and $g_{\rho\pi\pi}$ from the mass term $\chi^2 a {\cal L}_V$ \cite{Bando:1984ej,Bando:1987br}:
\begin{equation}
g_{\rho}=g F_\rho =2 F_\pi^2 g_{\rho\pi\pi} =2 v^2 g_{\rho\pi\pi}\,,
\end{equation}
and its extension to $W/Z-\rho$ mixing strength (low energy theorem of HLS: Proof in Ref. \cite{Harada:1993jk}) should be intact 
even when the mass term becomes dimension 4 with the extra factor $\chi^2$ which introduces additional symmetry, the scale symmetry). 
As usual in the Higgs mechanism, the gauge bosons of ${\rm gauged-}H_{\rm global} (\subset 
{\rm gauged-}G_{\rm global}$) get mixed with the gauge bosons of HLS, leaving  only the gauge bosons of the unbroken diagonal subgroup $({\rm gauged-}H)=H_{\rm local} + ({\rm gauged-}H_{\rm global})$ be massless after mass diagonalization.

     Again note that {\it the mass terms including the couplings of all the SM particles except for the Higgs  mass term $V(\phi)$ 
     are dimension 4 operators and thus are scale-invariant}.
 
\section{Dynamical Generation of HLS Gauge Boson I: Lesson from the Dynamical Higgs in the NJL Model \cite{Yamawaki:2015tmu} }

Before discussing the dynamical generation of the HLS gauge boson, kinetic term and Yang-Mills self-couplings as well,  here we 
 recapitulate   the well-known formulation \cite{Eguchi:1976iz,Bardeen:1989ds} to show  the dynamical generation of the composite Higgs model, kinetic term and quartic self coupling,
  based on the strong coupling phase $G>G_{\rm cr}\ne 0$ 
in the NJL  model. 
The NJL Lagrangian for the $N_C-$component 2-flavored fermion $\psi$ takes the form:
\begin{eqnarray}
{\cal L}_{\rm NJL} = \bar \psi i\gamma^\mu\partial_\mu \psi + \frac{G}{2} \left[ (\bar \psi \psi)^2 +(\bar \psi i \gamma_5 \tau^a \psi)^2\right]\,.
\end{eqnarray}
We can add the auxiliary field term which does not change the physics (Gaussian term trivially integrated out in terms of the path integral) :
\begin{eqnarray}
{\cal L}_{\rm aux}= - \frac{1}{2G} \left(\hat \sigma- G \bar \psi \psi\right)^2 - \frac{1}{2 G} \left(\hat \pi^a - G \bar \psi i \gamma_5 \tau^a \psi\right)^2\,. 
\end{eqnarray}
This in fact yields zero ${\cal L}_{aux}=0$ when equation of motion  $\hat\sigma=G \bar \psi \psi$ and $\hat \pi^a=G \bar \psi i \gamma_5 \tau^a \psi$ are used. These auxiliary field terms just correspond to the HLS term  in Eq. (\ref{HiggsHLS}) which also yields zero when the equation of motion for $\rho_\mu$ is used.
Then the resultant Lagrangian read
\begin{eqnarray}
{\cal L}_{\rm NJL}+{\cal L}_{\rm aux}= \bar \psi \left(i\gamma^\mu\partial_\mu  +\hat \sigma +i\gamma_5 \tau^a \hat \pi_a \right)\psi -\frac{1}{2}\frac{1}{G} \left({\hat \sigma}^2 +{\hat \pi_a}^2\right)\,,
\label{NJL}
\end{eqnarray}
where we can again check that 
the   equations of motion of the auxiliary fields  $\hat \sigma \sim G \bar \psi \psi$ and ${\hat \pi}^a \sim G \bar \psi i\gamma_5 \tau^a\psi$ are plugged back in the Lagrangian 
resulting in the original Lagrangian.

However at quantum level, a ``miracle'' takes place\footnote{Actually, it is not an miracle  nor magic, since the following formulation yields  precisely the same 
result as the traditional  direct large $N_C$ computation with gap equation and Bethe-Salpeter
equation,  giving in fact  the bound state Higgs dynamically generated.
}: in the large $N_C$ limit ($N_C \rightarrow \infty$ with $N_C G\ne 0$ fixed), we 
may integrate the loop contribution from the cutoff scale $\Lambda$ down to some infrared scale $\mu$ to have dynamical generated kinetic term and quartic coupling of $(\hat \sigma,\hat \pi^a)$ 
in the sense of the Wilsonian renormalization-group as 
\begin{eqnarray}
{\cal L}_{\rm induced}&=&\frac{1}{2} Z_\phi \left[(\partial_\mu \hat \sigma)^2 +(\partial_\mu \hat \pi)^2\right] +
\frac{1}{4} Z_\phi \left[ (\hat \sigma)^2+ (\hat \pi^2)\right]^2\,, \\
 Z_\phi&=& \frac{N_C}{8\pi^2} \ln \frac{\Lambda^2}{\mu^2}\,,
\end{eqnarray}
together with the mass shift
\begin{equation}
{\cal L}_{\rm mass}=\frac{1}{2} \left(\frac{1}{G}-\frac{N_C}{4\pi^2}(\Lambda^2-\mu^2)\right) \left({\hat \sigma}^2 +{\hat \pi_a}^2\right)\,. %
\end{equation}

When we take $\mu \rightarrow \Lambda$, all the induced terms disappear and  we get back to the original bare Lagrangian Eq.(\ref{NJL}).
The condition $Z_\phi(\mu) \rightarrow 0$ for $\mu \rightarrow \Lambda$ is the so-called compositeness condition \cite{Bardeen:1989ds}.
After rescaling the induced kinetic term to the canonical one, $Z_\phi^{1/2} (\hat \sigma, \hat \pi^a) \rightarrow (\hat \sigma,\hat \pi^a)$,
 the quantum theory for $\hat \sigma$ and $\hat \pi$ sector  yields 
precisely the same form as the SM Higgs Eq.(\ref{sigma}) \cite{Eguchi:1976iz}, with
\begin{eqnarray}
\mu_0^2&=& \left(\frac{1}{G} - \frac{N_C}{4\pi^2} (\Lambda^2-\mu^2)\right) Z_\phi^{-1}
=- \lambda v^2 < 0 \,,\nonumber\\
\lambda&=& Z_\phi Z_\phi^{-2} = Z_\phi^{-1} = \left[\frac{N_C}{8\pi^2} \ln \frac{\Lambda^2}{\mu^2
}\right]^{-1} \,,
\label{tachyon}
\end{eqnarray}
where we have used the relation $F_\pi^2=v^2=-\mu_0^2/\lambda$ in Eq.(\ref{chicircle}).  The quadratic running in mass is understood to be renormalized  into the 
bare mass term $1/G$ so as to keep $\mu_0^2<0$ (spontaneous breaking of the chiral symmetry) as usually done by the gap equation. On the other hand,
there is no bare term of the quartic coupling (unless we introduce 8-fermion operator), the induced quartic coupling has no renormalization such that $\lambda(\mu)$  has a Landau pole at $\mu=\Lambda$ in the SM language.\footnote{
There of course exists a conformal/BPS limit of the NJL model written in the form of the SM Higgs Lagrangian. See Ref.\cite{Yamawaki:2015tmu}. 
}

\section{Dynamical Generation of HLS Gauge Boson II: Standard Model Rho Meson \cite{Matsuzaki:2016iyq} }
 Now we discuss \cite{Matsuzaki:2016iyq}  that the kinetic term of $\rho_\mu$ is generated dynamically by the quantum loop \cite{Bando:1987br,Harada:2003jx}, in the same sense as that of the dynamical generation of the 
 kinetic term (and the quartic coupling as well) of the composite Higgs in the NJL model, which is an auxiliary field at the tree level or at composite scale \cite{Eguchi:1976iz}.
 
 In order to discuss the off-shell $\rho$ (in space-like momentum) relevant to the skyrmion stabilization to be discussed in section 6, we adopt the background field gauge as in Ref.\cite{Harada:2003jx}
 (where Feynman diagrams are explicitly given).
 The relevant diagrams for the two-point function of the SM rho at one loop are given in Fig.10 of Ref.\cite{Harada:2003jx}. 
 By integrating out the high frequency  modes  from the cutoff scale $\Lambda$ (composite scale) to the scale $\mu$ in the Wilsonian sense, we have
  the dynamically generated kinetic term as given in Eq.(4.194) of  Ref.\cite{Harada:2003jx} ($N_f=2$ in the case at hand):\footnote{There is a discontinuity 
  between  the results for $\mu \gg M_\rho$ and $\mu \ll M_\rho$ near $\mu \sim M_\rho$, which is an artifact  of disregarding the finite parts of the loop integral \cite{Harada:2003jx},
  and irrelevant to the discussions here.
    }
\begin{eqnarray}
{\cal L}^{(\rho)}_{\rm kinetic}&=& - \frac{1}{2g(\mu)^2} \,  {\rm tr} \rho_{\mu\nu}^2 \,, 
\label{kinetic} \\
 \frac{1}{g(\mu)^2} &=&\frac{1}{(4\pi)^2}\, \frac{a^2}{12}\, \ln \frac{\Lambda^2}{\mu^2} \quad \left(\mu^2 \ll M_\rho^2\right)\,, \nonumber \\
 &=& \frac{1}{(4\pi)^2}\, \left(\frac{a^2}{12}+ \left(-\frac{22}{3}+\frac{1}{12}\right)\right)\, \ln \frac{\Lambda^2}{\mu^2}  \quad  \left(\mu^2 \gg M_\rho^2\right) \,. 
 \label{RGErho}
 \end{eqnarray}

  For $\mu^2\ll M^2_\rho$ \footnote{Rescaling the kinetic term to the canonical one by  $\rho_\mu \rightarrow g \rho_\mu$, we have the off-shell mass $M_\rho^2(\mu) = g^2(\mu) F_\rho^2=a g^2(\mu) v^2$ which 
  behaves as $M_\rho(\mu) \rightarrow \infty (0)$ as $\mu \rightarrow \Lambda (0)$.  We have defined $M_\rho$ as $M^2_\rho\equiv  M^2_\rho(\mu=M_\rho) =g^2(\mu=M_\rho) F_\rho^2$.  } ,
  we have only the loop contribution of $\pi$ (longitudinal $W/Z$ when the electroweak gauging switched on) with $g_{\rho\pi\pi}=a/2$, which is
  characterized by $a^2/12$ in front of the log in Eq.(\ref{RGErho}). The resultant kinetic term includes the self-couplings of the SM rho as it should (in a way consistent with the gauge invariance), similarly to the composite Higgs self coupling as well as
 the kinetic term in the NJL model.   The kinetic term vanishes $\frac{1}{g^2(\mu)} \rightarrow 0$ for $\mu\rightarrow \Lambda$, in a way similar to composite Higgs in the NJL model \cite{Eguchi:1976iz}.  
 This is the simplest case when  $M_\rho ={\cal O} (\Lambda)$.

 On the other hand, for $\mu^2\gg M^2_\rho$, there arise additional contributions: the  gauge loop from the dynamically generated SM rho self coupling, with the usual factor, $-22/3$,
  together with $+1/12$ from the loop of the would-be NG boson ${\check \rho}$  (the longitudinal 
  SM rho) having the $\rho$ coupling $1/2$,   adds up to the characteristic factor $(a^2-87)/12$ instead of $a^2/12$.
   For the gauge coupling $g(\mu)^2$ to  have a Landau pole at $\mu=\Lambda$ in conformity of
 the compositeness condition, we would need  
 \begin{equation}
a>\sqrt{78}\simeq 9.3\quad \left(M_\rho\ll \Lambda\right)\,.
\end{equation}

Although Eq.(\ref{RGErho})  is similar to that of the  QCD rho meson,
an outstanding difference is that the QCD rho kinetic term is already generated by the underlying QCD and hence the standard QCD value \cite{Bando:1984ej,Bando:1987br},  $a=2$, 
has no problem about the HLS framework of the QCD rho meson, while in the SM rho case for  $M_\rho^2\ll \Lambda^2 $ we generally need a large value of $a$
 to get  the kinetic term to be dynamically generated
by the SM dynamics alone without recourse to the UV completion. 

Of course there is a possibility that even in the SM rho meson  the bare kinetic term may be provided by some underlying theory beyond the SM, such as the walking technicolor \cite{Yamawaki:1985zg}
whose low energy effective theory is  the scale-invariant HLS model \cite{Kurachi:2014qma} similar to the SM Higgs Lagrangian as discussed before.
In such a case,  a large $a$ would not be needed as in the QCD rho meson. 
 
In the next section we shall only discuss the simplest scenario $M_\rho={\cal O} (\Lambda)$ without such a UV completion as a benchmark for the SM skyrmion for the dark matter.

By  the loop effects, the $M_\rho^2$ would develop (potentially large) imaginary parts  in the time-like region for decaying to the $\pi\pi$ ($W_L W_L, W_L Z_L$ when
 the electroweak couplings switched on)  if $M_\rho> 2 M_{W/Z}$. However, this would not affect the skyrmion physics which is relevant to the
 space-like $\rho$.    Thus we will confine ourself to the kinetic term hereafter, although it would give rise to interesting collider physics if $M_\rho={\cal O}$ (TeV).
   
\section{SM Skyrmion as a Dark Matter, ``Dark SM Skyrmion'' \cite{Matsuzaki:2016iyq}}

Now that we have discussed the dynamical generation of the kinetic term of the SM rho meson, we can discuss the skyrmion living inside the SM Higgs Lagrangian in the form of
Eq.(\ref{qHiggsHLS}), which does have a skyrmion stabilized by the SM rho meson as a viable candidate for the dark matter, ``dark SM skyrmion (DSMS)'' denoted as $X_s$. \cite{Matsuzaki:2016iyq}. 
 
 It is in fact well known in the hadron physics \cite{Igarashi:1985et} that the QCD rho meson stabilizes the skyrmion (nucleon) with $I=J=1/2, 3/2, \cdots$, with a numerical result 
 giving
 rise to  
 the  skyrmion mass somewhat smaller than that in the original Skyrme model. 
 It was further shown \cite{Igarashi:1985et} that the Skyrme term may be regarded as the heavy mass limit of the HLS rho meson, $M_\rho^2=a g^2 v^2\gg v^2 $,  such that
$a\rightarrow \infty,\,\, g= {\rm constant}$, where 
the rho field configuration in the term  $\chi^2 a {\cal L}= a \chi^2 (\rho_\mu - {\alpha}_{\mu,||})^2$ in Eq.(\ref{qHiggsHLS}) is restricted to 
the auxiliary field configuration as a composite of the  $\pi$ in the nonlinear base: 
\begin{eqnarray}
 \rho_\mu \rightarrow {\alpha}_{\mu||} = \frac{1}{2i} \left(\partial_\mu \xi_{R}(x) \cdot \xi_{R}^\dagger(x)+ \partial_\mu \xi_{L}(x) \cdot \xi_{L}^\dagger(x) \right) 
 \nonumber \\  
 \left(a\rightarrow \infty,\,\, g= {\rm constant}\right)\,,
\end{eqnarray}
and hence 
\begin{eqnarray}
\rho_{\mu\nu} & \rightarrow& \rho_{\mu\nu}|_{\rho_\mu={\alpha}_{\mu||}} = i  [\hat\alpha_{\mu \perp}, \hat{\alpha}_{\nu \perp}]\,,
\quad \hat\alpha_{\mu \perp}=\frac{\xi_L (\partial_\mu U) \xi^\dagger_R}{2i}\,,\\
{\cal L}^{(\rho)}_{\rm kinetic}(\rho_\mu) 
&\rightarrow& 
- \frac{1}{2 g^2} {\rm tr} \left(  i  [\hat\alpha_{\mu \perp}, \hat{\alpha}_{\nu \perp}]    \right)^2
= \frac{1}{32 g^2} {\rm tr}[
 [ \partial_\mu U U^\dag, \partial_\nu UU^\dag ]^2
]
\,, \label{Skyrme-term}
\end{eqnarray} 
namely, the  kinetic term becomes
precisely equal to the Skyrme term with $e=g$ in this limit. 

It has also been shown \cite{Park:2003sd,Ma:2013ela,Kitano:2016ooc} that  including the scalar meson (corresponding to the SM
Higgs as a dilaton in our case) does not invalidate the skyrmion, but rather makes the skyrmion mass lighter. 
Particularly the skyrmion in QCD was discussed  \cite{Park:2003sd} in essentially the same form as our scale-invariant HLS form of the SM Higgs Lagrangian Eq.(\ref{qHiggsHLS}) 
except for  the potential term $V(\phi)$ (dimension 4 instead of our case with dimension 2) and adopted values of the parameters $F_\pi, F_\phi,a,g$, etc.. 

Thus the dark SM skyrmion (DSMS) $X_s$  emerges as a soliton solution from the Lagrangian Eq.(\ref{qHiggsHLS}) 
to be a topological bosonic matter carrying the topological number, which we call $U(1)_{X_s}$.\cite{Matsuzaki:2016iyq}
The $U(1)_{X_s}$ symmetry protects the decay of the DSMS 
($X_s$) completely, so the $X_s$ can be a dark matter candidate. Here we consider a
complex scalar DSMS with $I=J=0$. 

The DSMS is essentially generated by the scale-invariant part of Eq.(\ref{qHiggsHLS}) and hence 
its coupling is dictated by the nonlinear realization of the scale symmetry. In the low energy limit
$q^2 \ll v^2 \simeq (246 \, {\rm GeV})^2$ 
{\it the DSMS ($X_s$)  coupling to the SM Higgs $\phi$ as a pseudo dilaton is  
unambiguously determined by the low-energy theorem of the scale symmetry} \cite{Carruthers:1971vz}, 
as described in Eq.(\ref{LET}) \cite{Matsuzaki:2016iyq}
\begin{equation}
g_{\phi X_s X_s}=\frac{2 M^2_{X_s}}{ v}=2 v \lambda_{X_s}\,,\quad \lambda_{X_s}=\frac{M_{X_s}^2}{v^2} 
\label{DSMScoupling}
\end{equation}
which is relevant to the dark matter detection experiments for 
weakly-interacting massive-particle (WIMP) such as the LUX experiments~\cite{Akerib:2016vxi}.

 Through the Higgs $(\phi)$ exchange at zero-momentum transfer  as given in Eq.(\ref{DSMScoupling}), the relevant spin-independent (SI) elastic scattering cross section of the DSMS  
$X_s$ with the target nucleus, $({\rm Xe}: Z=54, A=131.293, u=0.931\,{\rm GeV})$  
per nucleon $(N=p,n)$,
can be calculated as 
\begin{eqnarray} 
&& 
\sigma_{\rm SI}^{\rm elastic/nucleon} (X_s N \to X_s N) 
= 
\frac{\lambda_{X_s}^2}{\pi M_\phi^4}  
\nonumber \\  
&& 
\times 
\left[ \frac{Z \cdot m_*(p, X_s) g_{\phi pp} + (A-Z) \cdot m_*(n,X_s) 
g_{\phi nn} }{A} 
 \right]^2
\,, \nonumber \\ 
\label{SI} 
\end{eqnarray} 
where $ m_*(N,X_s) = \frac{M_{X_s} m_N}{M_{X_s} + m_N} $ and  
$ g_{\phi pp(nn)} 
= 
\sum_q \sigma_q^{(p(n))}/v  
\simeq 
0.248(0.254)\,{\rm GeV}/v
$~\cite{Ohki:2008ff,Hisano:2015rsa}.

  \begin{figure}[t]
\begin{center}
   \includegraphics[scale=0.55]{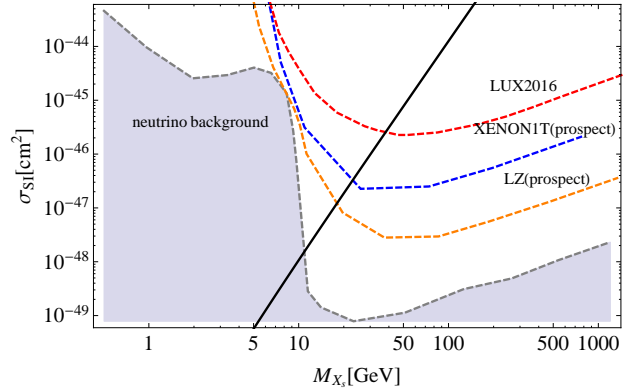} 
\caption{ 
The spin-independent elastic scattering cross section of the DSMS  
$X_s$ per nucleon as 
a function of the mass $M_{X_s}$ in units of ${\rm cm}^2$ (solid curve). 
The most stringent constraint at present from the latest 
LUX2016 experiment~\cite{Akerib:2016vxi} are shown 
and also shown are the projected experiments with the xenon target by the end of this 
decade~\cite{Feng:2014uja}.  
The gray domain, surrounded by the dashed curve on the bottom, 
stands for 
the atmospheric and astrophysical neutrino background~\cite{Billard:2013qya}.   
}
\label{mB-sigma}
\end{center} 
 \end{figure}

 Using $m_{p(n)} \simeq $ 938(940) MeV together with 
the electroweak scale $v \simeq 246$ GeV and the Higgs mass $M_\phi \simeq 125$ GeV, 
we can
numerically evaluate the cross section as a function of the DSMS 
mass $M_{X_s}$.   
We find \cite{Matsuzaki:2016iyq} (See Fig.~\ref{mB-sigma})
that the currently strongest exclusion limit from the 
latest LUX2016 experiment~\cite{Akerib:2016vxi} 
 implies 
\begin{equation} 
 M_{X_s} \lesssim 38 \,{\rm GeV} 
 \,.  \label{mB:limit:LUX}
\end{equation} 

Note that we have the {\it upper bound instead of the lower bound} in contrast to conventional WIMP models due to the characteristic dilatonic coupling proportional to $M_{X_s}^2$ as in Eq. (\ref{DSMScoupling}).

\begin{figure}[htbp]
\begin{center}
   \includegraphics[scale=0.45]{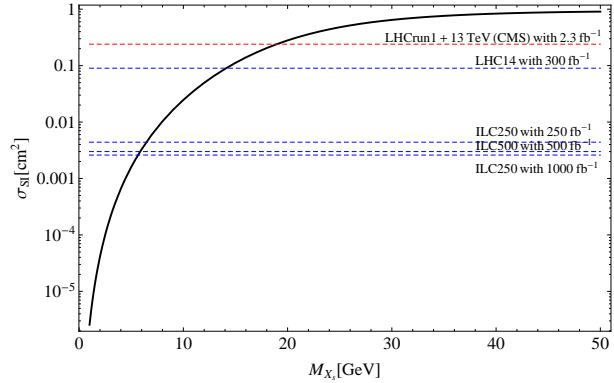} 
\caption{ 
The branching fraction of the Higgs decaying to the DSMS pair as 
a function of the mass $M_{X_s}$ (solid curve)
compared with the most stringent (95\% C.L.) 
constraint at present from the  
LHC run I combined with 
the early stage of the 13 TeV run  
reported by the CMS group~\cite{CMS:2016rfr}. Also plotted are 
the expected 95\% C.L. limits in the projected collider experiments including 
the 14 TeV LHC and ILC~\cite{Baer:2013cma}.}  
\label{mB-BR}
\end{center} 
 \end{figure}

Since the LUX2016 limit in 
Eq.(\ref{mB:limit:LUX}) implies $M_{X_s} < M_\phi/2 \simeq 63\, {\rm GeV}$,  further constraint on the mass of the
DSMS will be placed 
through the Higgs invisible decay limit 
at collider experiments.  
{\it The on-shell coupling of the SM Higgs as  
a pseudo dilaton to the  
 $X_s \bar{X}_s$, relevant to 
the invisible decay process, should be the same as 
that determined by the low-energy theorem, i.e., $q \sim M_\phi \ll v$ 
in Eq.(\ref{DSMScoupling}), as it is true for the  couplings of the SM Higgs as a pseudo dilaton
 to all the SM particles as mentioned before.}   
The partial decay width of the Higgs $\phi$ to the $X_s \bar{X}_s$ 
is thus unambiguously computed as 
\begin{equation}
 \Gamma(\phi \to X_s \bar{X}_s) 
= \frac{\lambda_{X_s}^2 v^2}{4 \pi M_\phi} 
\sqrt{1 - \frac{4 M_{X_s}^2}{M_\phi^2}}
\,. 
\end{equation}
The branching ratio is then constructed as 
$ 
{\rm Br}[\phi \to X_s \bar{X}_s] 
=
\Gamma(\phi \to X_s \bar{X}_s)/\Gamma_\phi^{\rm tot} 
= 
\Gamma(\phi \to X_s\bar{X}_s)/
[\Gamma_\phi^{\rm SM} + \Gamma(\phi \to X_s \bar{X}_s)]
$,  
with the total SM Higgs width (without the $X_s \bar{X}_s$ decay mode) 
$\Gamma_\phi^{\rm SM}\simeq 4.1$ MeV at the mass of 125 GeV~\cite{tot:higgs}. 
The currently most stringent upper limit on the Higgs invisible decay 
has been set by the CMS Collaboration combined with 
the run II data set with the luminosity of 
$2.3\,{\rm fb}^{-1}$~\cite{CMS:2016rfr}. 
Figure~\ref{mB-BR} shows the exclusion limit on the $X_s$ mass 
at 95\% C.L., 
 ${\rm Br}_{\rm invisible} \lesssim 0.2$~\cite{CMS:2016rfr}. 
 From the figure, we find \cite{Matsuzaki:2016iyq} 
\begin{equation} 
 M_{X_s} \lesssim 18 \, {\rm GeV}\,.
\label{mB:limit:BR}
\end{equation}
In Fig.~\ref{mB-BR} the future prospected 95\% C.L. limits 
in the LHC and ILC experiments~\cite{Baer:2013cma} are also shown.

We should emphasize  that such a characteristic mass range of the dark matter candidate is a salient feature of the scale symmetry
of the DSMS coupled to the SM Higgs as a pseudo dilaton as the low energy theorem of the scale symmetry. 
\\

Now we come to the relic abundance of the DSMS in the thermal history of the Universe. \cite{Matsuzaki:2016iyq}
In the thermal history of the universe 
the DSMS emerges after the electroweak phase transition at the temperature 
$T = {\cal O}(v)$. 
Below the freeze-out temperature $T<T_f$ ($x=M_{X_s}/T >x_f =M_{X_s}/T_f\simeq 20$),
 the DSMS number density evolves merely according to 
the adiabatic expansion of the universe, 
and the DSMS is cooled down  
to become a cold dark matter just like WIMPs,  
with the relic abundance observed in the Universe today. 
Such a relic abundance can be estimated by the standard 
procedure, so-called the freeze-out thermal relic~\cite{KT}: 
\begin{equation}
 \Omega_{X_s} h^2 
= 
\frac{2 \times (1.07 \times 10^9)x_f}
{g_*(T_f)^{1/2} M_{\rm Pl} {\rm GeV} \langle 
\sigma_{\rm ann} v_{\rm rel} \rangle} \,,
\end{equation}
where $M_{\rm pl}$ stands for the Planck mass scale $\simeq 10^{19}$ GeV, 
$\langle \sigma_{\rm ann} v_{\rm rel} \rangle $ 
is the thermal average of 
the annihilation cross section  
times the relative velocity of $X_s \bar{X}_s$, 
$v_{\rm rel}$,  
and $g_*(T_f)$ denotes the effective degrees of freedom for relativistic particles at $T=T_f$. 
The prefactor 2 comes from counting both $X_s$-particle 
and $\bar{X}_s$-anti-particle
 present today. 
The freeze out temperature $T_f$ can be determined by 
$ 
x_f = \ln [2 \times 0.038 \times [g_*(T_f) x_f]^{-1/2} M_{\rm pl}\cdot M_{X_s} \cdot 
\langle \sigma_{\rm ann} v_{\rm rel} \rangle  ] 
$. 

In evaluating the cross section, 
we expect that, until the freeze out time, 
the $X_s \bar{X_s}$ annihilation into the $U(1)_{X_s}$ current    
 will predominantly take place. 
Here  we note that 
the DSMS is 
a soliton, {\it an 
extended particle with 
a finite radius}.  
Such an annihilation process  
can be viewed as the classical $X_s \bar{X}_s$ collision 
with the $U(1)_{X_s}$ charge radius    
$R_{_{X_s}}=(\langle r_{_{X_s}}^2 \rangle_{_{X_s}})^{1/2}$, 
so 
\begin{equation}
\langle \sigma_{\rm ann} v_{\rm rel} \rangle = {\cal O}(\pi\, R^2_{_{X_s}})={\cal O}(\pi \cdot \langle r_{_{X_s}}^2 \rangle_{_{X_s}}
)
\end{equation}  
(Similar observation was made 
in Ref.~\cite{Kitano:2016ooc}.)  
Thus we need to evaluate 
the size of $\langle r^2_{_{X_s}} \rangle_{_{X_s}}
$,  in the standard skyrmion calculation with a scalar meson in the literature \cite{Park:2003sd,Ma:2013ela,Kitano:2016ooc}.
  
To get a rough idea of the DSMS as a dark matter, we here discuss\cite{Matsuzaki:2016iyq}  the simplest case, the heavy rho  (rho decoupled) limit mentioned above,  where the rho kinetic term may be replaced  by the Skyrme term Eq.(\ref{Skyrme-term}).
 The Lagrangian Eq.(\ref{qHiggsHLS}) reads:
 \begin{eqnarray}
{\cal L}_{\rm Higgs-Skyrme}&=& \chi^2(x) \cdot \left(
\frac{1}{2} \left(\partial_\mu \phi\right)^2 +
 \frac{v^2}{4}\cdot {\rm tr} \left(\partial_\mu U \partial^\mu U^\dagger \right) \right) -V(\phi) \nonumber \\
&+&  \frac{1}{32 g^2} {\rm tr}[
 [ \partial_\mu U U^\dag, \partial_\nu UU^\dag ]^2
] 
 \,,
 \label{qHiggsSkyrme}
 \end{eqnarray}
 which keeps the  scale-invariance of  Eq.(\ref{qHiggsHLS}), except for the potential $V(\phi)$.
 
 Then the skyrmion system is 
 essentially the same as the one  
analyzed in Ref.~\cite{Kitano:2016ooc}  as to the soliton solution and numerical results 
 in spite of the lack of scale symmetry in their Lagrangian and the associated  
scale-non-invariant form of the Higgs profile.
Using the standard spherically symmetric hedgehog ansatz, $U(\vec{x})=\exp(i \vec{\tau} \vec{\hat{r}} \, F(\vec{r}))$, 
we have \cite{Matsuzaki:2016iyq}
\begin{eqnarray}
M_{X_s}
&=&
\frac{2\pi v}{g} \int_0^\infty dr
 \left[
  \chi^2(r) \left(r^2 F^\prime(r)^2 +2 \sin^2(F(r))\right)  \right.\nonumber\\
&+& \left . \sin^2(F(r))\left(\frac{\sin^2(F(r))}{r^2} +2 F^\prime(r)^2\right)\right.\nonumber\\
&+& \left.
\chi^2(r) r^2 \phi^\prime(r)^2 +\frac{M_\phi^2}{4g^2 v^2} r^2\left(\chi(r)^4-2\chi(r)^2+1\right)
\right]
\nonumber\\
 &\simeq& 35 \, \frac{v}{g}\nonumber\,,\\
\langle r_{_{X_s}}^2 \rangle_{_{X_s}} 
&=& \int_0^\infty  dr \, r^2 \left(4\pi r^2\right) J^0_{X_s}=\frac{2}{\pi}\int_0^\infty
dr\, r^2 \sin^2(F(r)) F^\prime(r) \nonumber \\
&\simeq& 4.4\, \frac{1}{g^2 v^2}  \,,
\end{eqnarray}
for $g \gtrsim 478 (\gg 1)$ 
with $M_{X_s} \lesssim 18$ GeV in Eq.(\ref{mB:limit:BR}),
\footnote{
The equation of motion of $F(r)$ and $\phi(r)$ are given by 
$- \left[\chi^2 r^2+ 2\sin^2 F\right] \cdot F^{\prime\prime}=
\chi^2\left(-\sin(2 F)+2 r F^\prime\right) + 2 \chi \chi^\prime r^2 F^\prime -\sin^2F\sin(2F)/r^2+\sin(2F)(F^\prime)^2$ and
$\phi^{\prime\prime}=(F^\prime)^2+ 2 \sin^2(F)/r^2-(\phi^\prime)^2- 2\phi^\prime/r +M_\phi^2/((2 g^2v^2) (\chi^2-1)$,
with the boundary conditions $F(0)=\pi\,, F(\infty)=0$ and $\phi^\prime(0)=0,\, \phi(\infty)=0$. The solution implies that
$\chi(0)\rightarrow0\,(\phi(0)\rightarrow -\infty)$, i.e., symmetry restoration at the origin, in the conformal/BPS limit Eq.(\ref{conformallimit}), similarly to
the BPS limit of the 't Hooft-Polyakov monopole in the Georgi-Glashow model \cite{Harvey:1996ur}.
}
where we have defined  a dimensionless parameter ${\tilde r}=gv r $  and $F(r)={\tilde F} ({\tilde r})$, and renamed  
$\tilde r \rightarrow r$ and $\tilde F \rightarrow F$  
in the final expression for notational convenience.
$ J^0_{X_s}$ is the topological $U(1)_{X_s}$ current is defined by
\begin{eqnarray}
J^\mu_{X_s}=\frac{1}{24\pi^2}\epsilon^{\mu\nu\rho\sigma} {\rm tr} \left(U^\dagger \partial_\nu U\cdot U^\dagger\partial_\rho U\cdot U^\dagger \partial_\sigma U\right)\,.
\end{eqnarray}
and we have taken the topological charge $Q_{X_s}=1$.

Combining all of them together into the formulas given 
above,  
we find 
\begin{equation}
\Omega_{X_s} h^2 = {\cal O}(0.1)
\end{equation}
 at $M_{X_s}=18$ GeV 
(and $T_f \simeq 1$ GeV), 
which is roughly consistent with the presently observed 
dark matter relic $\simeq 0.12$~\cite{Ade:2013zuv}~\cite{footnote:ADM}.   
\\

\section{Summary and Discussions}
We have discussed a nobel role of the hidden local symmetry (HLS) in the standard model (SM) Higgs Lagrangian.
The SM Higgs Lagrangian was shown to be cast into precisely the scale-invariant nonlinear sigma model, with the SM Higgs being the pseudo dilaton, which was 
further shown to be
gauge equivalent to the scale-invariant version of the HLS Lagrangian.  Then the dynamical gauge boson ``SM rho meson'' of the HLS stabilizes the skyrmion,
  ``dark Standard Model skyrmion (DSMS)'' $X_s$, a novel candidate for the dark matter without recourse to beyond the SM.

A salient feature of DSMS is the nonlinearly realized scale invariance of the whole dynamics, 
which unambiguously determines the couplings of the DSMS to the SM Higgs as a pseudo-dilaton 
in terms of the low energy theorem of the
spontaneously broken scale invariance. This imposes a definite constraint on the mass $M_{X_s} \lesssim 38$ GeV 
from the direct detection experiments LUX2016. 
With such a mass smaller than half of the SM Higgs mass,  we have further constraint 
$M_{X_s} \lesssim 18$ GeV from the SM Higgs invisible decay data 
also definitely constrained by the low-energy theorem. Based on this salient constraint we have discussed that the DSMS in a benchmark case 
of the heavy SM rho mass limit (scale-invariant Skyrme model limit)  is
 roughly consistent with the relic abundance  $\Omega_{X_s} h^2 \simeq 0.12$ in view of the extended size of the soliton.

If this scenario within the SM is established, then we should look for the beyond-SM physics somewhere other than the dark matter which is currently regarded as the biggest motivation for going beyond the SM.

The discussion here is straightforwardly applied to some UV completion, in which case the bare kinetic term of the HLS gauge boson is
already generated by the underlying theory, the situation similar to the QCD rho meson and the QCD skyrmion stabilized by the
QCD rho. A typical example for such a UV completion is 
the walking technicolor having the approximate scale symmetry and thus the technidilaton \cite{Yamawaki:1985zg}, which is described 
by essentially the same type of the scale-invariant nonlinear chiral Lagrangian as the present theory (see e.g., \cite{Fukano:2015zua}), with
technidilaton identified as the 125 GeV Higgs, though having somewhat larger decay constant  $F_\phi> v$. With the HLS gauge bosons being the technirho and  the skyrmion the technibaryon and largeness of $F_\phi>v $, the mass upper bound of the dark matter technibaryon from the
LUX constraint (see Fig. 1) shifts to the larger mass range, say $F_\phi >v$ for a typical walking technicolor case $F_\phi \simeq 5 v$\cite{Matsuzaki:2015sya}
, so that the Higgs invisible decay constraint becomes irrelevant. Also the technibaryon usually with mass on order of TeV can be
light when including   the effects of the scalar meson (pseudo dilaton in our case)  on the skyrmion solution as already observed in the literature \cite{Park:2003sd,Ma:2013ela,Kitano:2016ooc}.

We also comment \cite{Matsuzaki:2016iyq} on the possibility for the DSMS as an asymmetric dark matter,
which generates the current relic abundance of the DSMS through the electroweak sphaleron process collaborating with  
both the DSMS and the SM Higgs nonlinear dynamics of HLS  with a ``HLS sphaleron''.

In summary we have discussed a novel possibility that the HLS together with skyrmion, another pet idea of Gerry \cite{Zahed:1986qz}, will give a great impact in the sense somewhat other than that Gerry was anticipating but I hope he would certainly enjoy it with a nice  smile.

\section{Acknowledgments}
I would like to thank Shinya Matsuzaki and Hiroshi Ohki for collaborations and helpful discussions on the dark SM skyrmion. 
I also thank Mannque Rho for stimulating correspondence on the scale symmetry and pseudo dilaton in the gauge theory and the effective theory.

\bibliographystyle{ws-rv-van}
\bibliography{ws-rv-sample}

\end{document}